\documentstyle[12pt]{article}

\begin{document}

\begin{titlepage}

\begin{flushright}
IJS-TP-95/18\\
CMU-HEP-95-22 \\
NUHEP-TH-95-13\\
November 1995\\
\end{flushright}

\vspace{.5cm}

\begin{center}
{\Large \bf An effective model for
charmed meson semileptonic decays}
\vspace{1.5cm}

{\large \bf B. Bajc $^{a,b}$, S. Fajfer $^{a}$ and
R. J. Oakes $^{c}$}

\vspace{.5cm}

{\it a) J. Stefan Institute, University of Ljubljana,
61111 Ljubljana, Slovenia}

\vspace{.5cm}

{\it b)
Department of Physics, Carnegie Mellon University,
Pittsburgh, Pennsylvania 15213, U.S.A.}

\vspace{.5cm}

{\it c) Department of Physics and Astronomy
Northwestern University, Evanston, Il 60208
U.S.A.}

\vspace{1cm}

\end{center}

\centerline{\large \bf ABSTRACT}

\vspace{0.5cm}

We analyze charm meson semileptonic decays
using the measured ratios $\Gamma_L/\Gamma_T$
and $\Gamma_{+}/\Gamma_{-}$ from $D\to {\bar K}^{*}$
and the branching ratios for $D \to {\bar K}^{*}$ and
$D \to {\bar K}$. First we introduce the light vector
mesons in a model which combines the heavy quark
effective Lagrangian and the chiral perturbation approach.
We propose a method which predicts the behavior of the
form factors. Using the available experimental data we
determine the values of some model parameters and
reproduce the observed branching ratios for
$D_{s} \to \Phi$, $D_{s} \to (\eta + \eta^{\prime})$
and $D \to \pi$. We make predictions for the yet
unmeasured branching ratios and polarization observables.

\end{titlepage}

\centerline{\bf I. INTRODUCTION}

\vskip 1cm

There exist several theoretical calculations aiming to describe
$D \to V l \nu$ ($V$ is a light vector meson) or
$D \to P l \nu$ ($P$ is a light pseudoscalar meson) semileptonic
decays: relativistic or nonrelativistic
quark models \cite{stech}, \cite{isgur}, \cite{isgw},
\cite{gilman}, lattice calculations \cite{lubicz},
QCD sum rules \cite{ball} and a few attempts to use
the heavy quark effective theory (HQET) \cite{casone},
\cite{castwo}, \cite{kamal}.
On the experimental side the following quantities
regarding semileptonic charm meson decays have
been measured: the branching ratios {\cal B},
$\Gamma_L/\Gamma_T$ and $\Gamma_+/\Gamma_-$ for
$D^{+}\to {\bar K}^{*0}$ and the branching ratios {\cal B}
for $D^{0}\to K^{*-}$, $D^{+}_{s} \to \Phi$,
$D^{0}\to K^{-}$, $D^+\to {\bar K}^0$,
$D^{+}_{s} \to (\eta + \eta^{\prime})$,
$D^{0} \to \pi^{-}$ and $D^{+} \to \pi^{0}$ \cite{pdg}.
The purpose of this paper is to
accommodate these  available experimental
data within a combination of
HQET and the chiral perturbation theory (CHPT)
description of the light meson sector.
Within this framework the HQET is valid at small
recoil momentum \cite{wisgur}, \cite{wise}
(a zero recoil momentum is
realized when the final and decaying meson states
have the same velocities). HQET can give definite
predictions for heavy to light ($D\to V$ or $D\to P$)
semileptonic decays in the kinematic region
with large momentum transfer to the lepton
pair, i.e large $q^2$. Unfortunately, it
cannot predict the $q^2$ dependence of the form factors.

The experimental data for the semileptonic decays
$D^0\to K^-$ \cite{anjosone}, $D^+\to \bar{K}^0$
\cite{anjostwo} and $D^+\to\bar{K}^{*0}$
\cite{anjosthree}, \cite{kodama} are unfortunately not
good enough to clearly determine the $q^2$
dependence of the form factors. Experimentally what is
known, apart from the branching ratios, are the form factors
at one kinematical point, {\it assuming} a pole-type
behaviour for all the form factors. The same assumption
is used also in many theoretical calculations, for
example in \cite{stech}, \cite{castwo}.
This assumption seems reasonable, but
within HQET the kinematic constraint on the form
factors at $q^{2}= 0 $ cannot be satisfyied
unless a special relation is imposed between
the pole masses and residues. Moreover,
it was shown using QCD sum rules \cite{ball} that the
form factors for axial currents exhibit a rather
flat $q^{2}$ dependence.

For these reasons, we will modify the Lagrangian for
heavy and light pseudoscalar and vector mesons
given by the HQET and chiral symmetry \cite{casone}.
Apart from the zero-recoil point we will still use
the same Feynman rules for the vertices in our
processes, but write down the complete propagator
also for heavy mesons, instead of using the
HQET propagator. In the region where the heavy
meson is nearly on-shell (the region where HQET
is applicable) the two prescription almost
perfectly overlap,
but due to a Feynman rule prescription for
the calculation of the form factors, there
are no inconsistencies at $q^2=0$. At the
same time this gives a natural explanation of the
pole-type form factors in the whole $q^2$ range
and an entirely consistent picture. It enables us
to determine which form factors have a
pole-type or a constant behaviour, confirming the
results of the QCD sum rules analysis \cite{ball}.

In order to show that such a simple prescription works,
we will calculate the decay widths in all measured charm
meson semileptonic decays. The model parameters will then
be determined by the experimental data. These parameters are
also important in the study of more complicated decays.

The paper is organized as follows: in Sec. II we will
first write down the already known Lagrangian for heavy
and light psedoscalar and vector mesons, given by the
requirements of HQET and chiral symmetry. In Sec. III
we will explain the behavior of the form factors
in decays $D \to V$  and $D \to P$. The free model
parameters will then be determined by comparing our
approach with experiment. Finally, a short summary
of the results will be given in Sec. IV.

\vskip 1cm

\centerline{\bf II. THE HQET AND CHPT LAGRANGIAN}

\vskip 1cm

\noindent
{\bf a) Strong Interactions}

\vskip 0.5cm

We incorporate in our Lagrangian
both the heavy flavour $SU(2)$ symmetry \cite{wisgur},
\cite{georgi} with the $SU(3)_L\times SU(3)_R$ chiral
symmetry, spontaneously broken to the diagonal
$SU(3)_V$ \cite{bando}, which can be used for the
description of heavy and light pseudoscalar and
vector mesons. A similar Lagrangian, but without the
light vector octet, was first introduced by Wise
\cite{wise}, Burdman and Donoghue \cite{burdman},
and Yan et al. \cite{yan}.
It was then generalized with the inclusion of light
vector mesons in \cite{casone}, \cite{kamal}, \cite{ourpaper}.

The light degrees of freedom are described by the
3$\times$3 Hermitian matrices

\begin{eqnarray}
\label{defpi}
\Pi = \pmatrix{
{\pi^0\over\sqrt{2}}+{\eta_8\over\sqrt{6}}+{\eta_0\over\sqrt{3}} &
\pi^+ & K^+ \cr
\pi^- & {-\pi^0\over\sqrt{2}}+{\eta_8\over\sqrt{6}}+
{\eta_0\over\sqrt{3}} & K^0 \cr
K^- & {\bar K^0} & -{2 \over \sqrt{6}}\eta_8+
{\eta_0\over\sqrt{3}} \cr}\;,
\end{eqnarray}

\noindent
and

\begin{eqnarray}
\label{defrho}
\rho_\mu = \pmatrix{
{\rho^0_\mu + \omega_\mu \over \sqrt{2}} & \rho^+_\mu & K^{*+}_\mu \cr
\rho^-_\mu & {-\rho^0_\mu + \omega_\mu \over \sqrt{2}} & K^{*0}_\mu \cr
K^{*-}_\mu & {\bar K^{*0}}_\mu & \Phi_\mu \cr}
\end{eqnarray}

\noindent
for the pseudoscalar and vector mesons, respectively.
The mass eigenstates are defined by
$\eta=\eta_8\cos{\theta_P}-\eta_0\sin{\theta_P}$ and
$\eta'=\eta_8\sin{\theta_P}+\eta_0\cos{\theta_P}$, where
$\theta_P=(-20\pm 5)^o$ \cite{pdg} is the $\eta-\eta'$
mixing angle.
The matrices (\ref{defpi}) and (\ref{defrho}) are
usually expressed through the combinations

\begin{eqnarray}
\label{defu}
u & = & \exp  ( \frac{i \Pi}{f} )\;,
\end{eqnarray}

\noindent
where $f$ is the pseudoscalar decay constant, and

\begin{eqnarray}
\label{defrhohat}
{\hat \rho}_\mu & = & i {g_V \over \sqrt{2}} \rho_\mu\;,
\end{eqnarray}

\noindent
where $g_V=5.9$ is given by the values of the
vector masses (we consider only the case of exact
vector dominance, see \cite{ourpaper}).

Introducing the vector and axial currents
${\cal V}_{\mu} =  \frac{1}{2} (u^{\dag}
\partial_{\mu} u + u \partial_{\mu}u^{\dag})$
and ${\cal A}_{\mu}  =  \frac{1}{2} (u^{\dag}
\partial_{\mu} u - u \partial_{\mu}u^{\dag})$
and the guage field tensor
$F_{\mu \nu} ({\hat \rho}) =
\partial_\mu {\hat \rho}_\nu -
\partial_\nu {\hat \rho}_\mu +
[{\hat \rho}_\mu,{\hat \rho}_\nu]$
the light meson part of the strong
Lagrangian can be written as

\begin{eqnarray}
\label{defllight}
{\cal L}_{light} = &-&{f^2 \over 2}
\{tr({\cal A}_\mu {\cal A}^\mu) +
2\, tr[({\cal V}_\mu - {\hat \rho}_\mu)^2]\}\nonumber\\
& + & {1 \over 2 g_V^2} tr[F_{\mu \nu}({\hat \rho})
F^{\mu \nu}({\hat \rho})]\;.
\end{eqnarray}

Both the heavy pseudoscalar and the heavy vector
mesons are incorporated in the $4\times 4$ matrix

\begin{eqnarray}
\label{defh}
H_a& = & \frac{1}{2} (1 + \!\!\not{\! v}) (D_{a\mu}^{*}
\gamma^{\mu} - D_{a} \gamma_{5})\;,
\end{eqnarray}

\noindent
where $a=1,2,3$ is the $SU(3)_V$ index of the light
flavours, and $D_{a\mu}^*$ and $D_{a}$ annihilate a
spin $1$ and spin $0$ heavy meson $c \bar{q}_a$ of
velocity $v$, respectively. They have a mass dimension
$3/2$ instead of the usual $1$, so that the Lagrangian
is explicitly mass independent in the heavy quark
limit $m_c\to\infty$. Defining

\begin{eqnarray}
\label{defhbar}
{\bar H}_{a} & = & \gamma^{0} H_{a}^{\dag} \gamma^{0} =
(D_{a\mu}^{* \dag} \gamma^{\mu} + D_{a}^{\dag} \gamma_{5})
\frac{1}{2} (1 + \!\!\not{\! v})\;,
\end{eqnarray}

\noindent
we can write the leading order strong  Lagrangian as

\begin{eqnarray}
\label{deflstrong}
{\cal L}_{even} & = & {\cal L}_{light} +
i Tr (H_{a} v_{\mu} (\partial^{\mu}+{\cal V}^{\mu})
{\bar H}_{a})\nonumber\\
& + &i g Tr [H_{b} \gamma_{\mu} \gamma_{5}
({\cal A}^{\mu})_{ba} {\bar H}_{a}]
 +  i \beta Tr [H_{b} v_{\mu} ({\cal V}^{\mu}
- {\hat \rho}^{\mu})_{ba} {\bar H}_{a}]\nonumber\\
& + &  {\beta^2 \over 4 f^2 }
Tr ({\bar H}_b H_a {\bar H}_a H_b)\;.
\end{eqnarray}

This Lagrangian contains two unknown parameters,
$g$ and $\beta$, which are not determined by symmetry
arguments, and must be determined empirically.
This is the most general even-parity Lagrangian
in leading order of the heavy quark mass
($m_Q\to\infty$) and chiral symmetry limit
($m_q\to 0$ and the minimal number of derivatives).

We will also need the odd-parity Lagrangian for the
heavy meson sector. The lowest order contribution
to this Lagrangian is given by

\begin{eqnarray}
\label{defoddheavy}
{\cal L}_{odd} & = & i {\lambda} Tr [H_{a}\sigma_{\mu \nu}
F^{\mu \nu} (\hat \rho)_{ab} {\bar H_{b}}]\;.
\end{eqnarray}

\noindent
The parameter $\lambda$ is free, but we know that
this term is of the order $1/\Lambda_\chi$ with
$\Lambda_\chi$ being the chiral perturbation theory
scale \cite{CG}.

\vskip 0.5cm

\noindent
{\bf b) Weak Interactions}

\vskip 0.5cm

For the semileptonic decays the weak Lagrangian
is given at the quark level
by the current-current Fermi interaction

\begin{equation}
\label{deflfermisl}
{\cal L}_{SL}^{eff}(\Delta c=\Delta s=1)
=-{G_F \over \sqrt{2}}
[ ({\bar l}u_l)^\mu ({\bar s'}c )_\mu ]\;,
\end{equation}

\noindent
where $G_F$ is the Fermi constant,
$({\bar \psi}_1\psi_2)^\mu\equiv
{\bar \psi}_1\gamma^\mu(1-\gamma^5)\psi_2$ and
$s'=s\cos{\theta_C}+d\sin{\theta_C}$, $\theta_C$
being the Cabibbo angle ($\sin{\theta_C}\approx 0.222$).

At the meson level we assume that the weak
current transforms as $({\bar 3}_L,1_R)$ under
chiral $SU(3)_L\times SU(3)_R$ and is linear in the heavy
meson field. The most general current can then be written as

\begin{equation}
\label{current}
J_\lambda=(D^{*\mu}{\hat A}_{\mu\lambda}+
D{\hat B}_\lambda) u^\dagger\;,
\end{equation}

\noindent
The leading order in the $1/M$ expansion is obtained
by demanding that the operators ${\hat A}$ and ${\hat B}$
do not act as derivatives on the heavy meson fields $D$
and $D^*$ \cite{schechter}.
We can generally expand ${\hat A}_{\mu\lambda}$
and ${\hat B}_\lambda$ in powers of the operators

\begin{eqnarray}
\label{operators}
\hat{\cal O}^{(1)}_\mu & = & {\hat\rho}_\mu-
                             {\cal V}_\mu\;,\nonumber\\
\hat{\cal O}^{(2)}_\mu & = & {\cal A}_\mu\;,\nonumber\\
\hat{\cal O}^{(3)}_\mu & = & \partial_\mu+{\cal V}_\mu\;,
\end{eqnarray}

\noindent
which are, together with the quark mass matrix insertion,
the basic operators in our model with the correct transformation
properties. Using the standard (chiral) power counting rules
\cite{weinberg},
it is easily shown, that the operators (\ref{operators})
count as order ${\cal O}(E)$, having each one
derivative or one (light) vector field, while the mass
matrix is of higher order, ${\cal O}(E^2)$. Expanding
${\hat A}$ and ${\hat B}$ to order ${\cal O}(E)$,
i.e. to the linear order in powers of the operators
(\ref{operators}), we get, recalling that $v_\mu D^{*\mu}=0$,
where $v$ is the heavy meson four-velocity,

\begin{equation}
\label{defa}
\hat{A}^{\mu\lambda}=A g^{\mu\lambda}+[
B_{1,i}g^{\mu\lambda}v^\alpha+
B_{2,i}g^{\mu\alpha}v^\lambda
+B_{3,i}i\epsilon^{\mu\lambda\alpha\beta}v_\beta]
\hat{\cal O}^{(i)}_\alpha+...\;,
\end{equation}

\begin{equation}
\label{defb}
\hat{B}^\lambda=C v^\lambda+
[D_{1,i}g^{\lambda\alpha}+
D_{2,i}v^\lambda v^\alpha]\hat{\cal O}^{(i)}_\alpha+...\;.
\end{equation}

It is clear that such an expansion is a {\it chiral}
expansion, i.e. in powers of energy $E$, and need {\it not}
be a heavy quark expansion in powers of $1/M$, as it is
sometimes assumed \cite{grinstein}.
The current (\ref{current})
together with (\ref{defa})-(\ref{defb})
is the most general one in our model to order
${\cal O}(M^0)$ in the heavy quark expansion and
to order ${\cal O}(E^0)$ and ${\cal O}(E)$ in
the chiral expansion.

The ${\cal O}(E^0)$ coefficients, $A$ and $C$, can be
expressed in terms of the heavy meson pseudoscalar and
vector decay constants $f_D$ and $f_{D*}$, which
are equal in the heavy quark limit \cite{wisgur}, while
no such relations exist between the coefficients
$B_{i,k}$ (\ref{defa}) and $D_{i,k}$ (\ref{defb}).
However, due to the relation $\hat{\cal O}^{(3)}_\mu u^\dagger=
-\hat{\cal O}^{(2)}_\mu u^\dagger$, only the operators
$\hat{\cal O}^{(1)}_\mu$ and $\hat{\cal O}^{(2)}_\mu$
are important at order ${\cal O}(E)$.

In our calculation of the $D$ meson semileptonic decays to
leading order in both $1/M$ and the chiral expansion we
will need only the current proportional to
$D$, $DP$, or $D^*$ at order ${\cal O}(E^0)$ and the
current proportional to $DV$ at order ${\cal O}(E)$.
Consequently, we can rewrite Eq. (\ref{current})
with (\ref{defa})-(\ref{defb}) as

\begin{eqnarray}
\label{j}
{J}_{a}^{\mu} = &\frac{1}{2}& i \alpha Tr [\gamma^{\mu}
(1 - \gamma_{5})H_{b}u_{ba}^{\dag}]\nonumber\\
&+& \alpha_{1}  Tr [\gamma_{5} H_{b} ({\hat \rho}^{\mu}
- {\cal V}^{\mu})_{bc} u_{ca}^{\dag}]\nonumber\\
&+&\alpha_{2} Tr[\gamma^{\mu}\gamma_{5} H_{b} v_{\alpha}
({\hat \rho}^{\alpha}-{\cal V}^{\alpha})_{bc}u_{ca}^{\dag}]+...\;,
\end{eqnarray}

\noindent
where $\alpha=f_D\sqrt{m_D}$ \cite{wise}. The $\alpha_1$ term was
first considered in \cite{casone}. We also include the $\alpha_2$
term, as we must,
since it is of the same order in the $1/M$ and chiral
expansion as the term proportional to $\alpha_1$.
We will also see, that the term proportional to
$\alpha_2$ is very important for the phenomenology
of the semileptonic decays and that it can not
be neglected. In the next Section we will determine
$\alpha_1$ and $\alpha_2$ from the experimental data on
semileptonic $D$ meson decays.

\vskip 1cm

\centerline{\bf III. FORM FACTORS FOR $D\to V/P$}

\vskip 1cm

The description of semileptonic decays is
known near the zero-recoil point, but for the calculation
of the branching ratios we need to extrapolate the
form factors to different kinematical regions,
defined by the square of the transfer momentum $q^2$.
This has been done using the QCD sum-rule analysis \cite{ball},
quark models \cite{stech}, \cite{isgur},
\cite{isgw}, \cite{gilman}, and
lattice calculations \cite{lubicz}. Within a Lagrangian
approach this is more difficult, since the form of
the interactions is known only in the heavy meson mass
limit near zero-recoil. Therefore it was assumed in
\cite{stech} and \cite{castwo}, that all form factors
are pole-type functions of $q^2$, but
with different pole-masses. With the known values of
form factors in the zero-recoil limit, given by the
HQET, the form factors in the whole kinematic region
then seem to be determined. However, this prescription
possesses some shortcomings, which can be seen as
follows: the $H\to V$ and $H\to P$ current matrix
elements can be quite generally parametrized as

\begin{eqnarray}
\label{defhv}
<V(\epsilon,p')|(V-A)^\mu|H(p)>=
-{2 V(q^2)\over m_H+m_V}
\epsilon^{\mu\nu\alpha\beta}\epsilon_\nu^* p_\alpha
{p'}_\beta \nonumber\\
-i \epsilon^*.q {2 m_V\over q^2}q_\mu A_0(q^2)
+i(m_H+m_V)(\epsilon_\mu^*-
{\epsilon^*.q\over q^2}q_\mu)A_1(q^2) \nonumber\\
-{i\epsilon^*.q\over m_H+m_V}((p+p')_\mu-
{m_H^2-m_V^2\over q^2}q_\mu)A_2(q^2)\;,
\end{eqnarray}

\noindent
and

\begin{eqnarray}
\label{defhp}
<P(p')|(V-A)_\mu|H(p)>&=&[(p+p')_\mu-
{m_H^2-m_P^2\over q^2}q_\mu]F_1(q^2)\nonumber\\
&+&{m_H^2-m_P^2\over q^2}q_\mu F_0(q^2)\;,
\end{eqnarray}

\noindent
where, $q=p-p'$ is the exchanged momentum. In order
that these matrix elements are finite at $q^2=0$,
the form factors must satisfy the relations

\begin{equation}
\label{relff}
A_0(0)+{m_H+m_V\over 2 m_V}A_1(0)-
{m_H-m_V\over 2 m_V}A_2(0)=0\;.
\end{equation}

\begin{equation}
\label{f1f0}
F_1(0)=F_0(0)\;.
\end{equation}

\noindent
But these equations cannot be satisfied by calculating the form
factors at zero recoil where $q^2=q^2_{max}=(m_H-m_{V(P)})^2$
and then extrapolating them to $q^2=0$
assuming a simple pole $q^2$ dependence, unless a
relation between the model parameters is imposed.
It is unreasonable to assume such a
relation, since the pole masses are taken from the
measured lowest lying resonances with the correct
quantum numbers and, therefore, are not free parameters.

The problem, therefore, is how to extrapolate the amplitude
from the zero recoil point to the rest of the
allowed kinematical region. We shall make a very simple,
physically motivated, assumption:
the vertices do not change significantly, while the
propagators of the off-shell heavy mesons are given by
the full propagators $1/(p^2-m^2)$ instead of the HQET
propagators $1/(2 m v.k)$. With these assumptions
we are able to incorporate the following features:

\noindent
i) almost exactly the HQET prediction at the maximum $q^2$;

\noindent
ii) a natural explanation for the pole-type
form factors when appropriate;

\noindent
iii) predictions of flat $q^2$ behaviour for the form factors
$A_1$ and $A_2$, which has been confirmed in the QCD sum-rule
analysis of \cite{ball}.

Our approach is different than in
\cite{casone}-\cite{castwo}, where a pole
dominance prescription for the $q^2$
dependence of the form factors was
assumed, as in the data analysis of the
semileptonic $D^0\to K^-$ \cite{anjosone},
$D^+\to\bar{K}^0$ \cite{anjostwo} and
$D^+\to\bar{K}^{*0}$ \cite{anjosthree}, \cite{kodama}.
In contrast, we calculate the form factors
directly from our Lagrangian. For the strongly
off-shell charm meson propagator we take the
complete expression $1/(q^2-m_D^2)$ rather
than the heavy meson limit $1/(2 m_D vk)$, where
$k=q-m_D v$ is the residual momentum (assuming
for the moment the degeneracy of the $D-D^*$
system). The difference between the two approaches
at $q^2_{max}$ is less than $25\%$. Also,
in our approach the pole structure $1/(1-q^2/m_D^2)$
of certain diagrams is a direct consequence
of the $D$ or $D^*$ full propagators. From the
other side, not all diagrams have intermediate
$D$ or $D^*$ mesons and then such a $q^2$
dependence is absent. Consequently,  we have a
very simple way to determine if a particular
form factor has a pole-type behaviour, a constant
behaviour, or some combination.

Finally, we include $SU(3)$ symmetry breaking by
using the phy\-si\-cal mass\-es and decay constants
shown in Table \ref{tabone}.
The decay constants for $\eta$ and $\eta'$ were taken from
\cite{etaetap}, for $D_s$ from \cite{cleo}, while for the other
$D$'s theoretical predictions were used \cite{casthree}.

\vskip 0.5cm

\noindent
{\bf Decays $D\to Vl\nu_l$}

\vskip 0.5cm

There are three possible Cabibbo allowed semileptonic
decays ($D^0\to K^{*-}$, $D^+\to\bar{K}^{*0}$ and
$D^{s+}\to \Phi$) and four possible Cabibbo suppressed
semileptonic decays ($D^0\to\rho^-$,
$D^+\to\rho^0$, $D^+\to\omega$ and $D_s^+\to K^{*0}$)
of the charmed mesons of the type $D\to V l \nu_l$.
The relevant form factors defined in (\ref{defhv}), calculated
in our model, are

\begin{eqnarray}
\label{enfirst}
{1\over K_{HV}} V(q^2)=&\Big[&2(m_H+m_V)
({m_{H'^*}\over m_H})^{1\over 2}
{m_{H'^*} \over q^2-m_{H'^*}^2} f_{H'^*}\lambda\Big]{g_V\over\sqrt{2}}\;,\\
{1\over K_{HV}} A_0(q^2)=&\Big[&{1\over m_V}
({m_{H'}\over m_H})^{1\over 2}
{q^2 \over q^2-m_{H'}^2}f_{H'}\beta\nonumber\\
&-&{\sqrt{m_H}\over
m_V}\alpha_1+{1\over 2}
({q^2+m_H^2-m_V^2\over m_H^2})
{\sqrt{m_H}\over m_V} \alpha_2\Big]{g_V\over\sqrt{2}}\;,\\
{1\over K_{HV}} A_1(q^2)=&\Big[&-{2\sqrt{m_H}\over m_H+m_V}
\alpha_1\Big]{g_V\over\sqrt{2}}\;,\\
\label{enlast}
{1\over K_{HV}} A_2(q^2)=&\Big[&-{m_H+m_V\over m_H\sqrt{m_H}}
\alpha_2\Big]{g_V\over\sqrt{2}}\;,
\end{eqnarray}

\noindent
where the pole mesons and the corresponding constants
$K_{HV}$ are given in Table \ref{tabtwo}.
It is convenient to introduce the helicity amplitudes
for the decay $H\to V l^+\nu_l$ as in \cite{ball}:

\begin{eqnarray}
\label{helampone}
H_\pm(y)=&+&(m_H+m_V)A_1(y)\mp
{2 m_H |\vec{p'}(y)|\over m_H+m_V} V(y)\;,\\
H_0(y)=&+&{m_H+m_V\over 2 m_H m_V
\sqrt{y}}[m_H^2(1-y)-m_V^2] A_1(y)\nonumber\\
\label{helamptwo}
&-&{2 m_H \vec{p'}^2(y)\over m_V (m_H+m_V)\sqrt{y}}A_2(y)\;,
\end{eqnarray}

\noindent
where

\begin{equation}
\label{defy}
y={q^2\over m_H^2}\;,
\end{equation}

\noindent
and

\begin{equation}
\label{defpp}
|\vec{p'}|^2={[m_H^2 (1-y)+m_V]^2\over 4 m_H^2}-m_V^2\;.
\end{equation}

In order to compare with experiment, we calculate the
decay rates for polarized final light vector mesons:

\begin{equation}
\label{defgammapol}
\Gamma_a={G_F^2 m_H^2\over 96 \pi^3}
\int_0^{y_m}dy y |\vec{p'}(y)||H_a(y)|^2\;,
\end{equation}

\noindent
where $a=+,-,0$ and

\begin{equation}
\label{defym}
y_m=(1-{m_V\over m_H})^2\;.
\end{equation}

\noindent
The transverse, longitudinal and total decay rates
are then trivially given by

\begin{eqnarray}
\label{defgammaone}
\Gamma_T&=&\Gamma_++\Gamma_-\;,\\
\label{defgammatwo}
\Gamma_L&=&\Gamma_0\;,\\
\label{dsefgammathree}
\Gamma&=&\Gamma_T+\Gamma_L\;.
\end{eqnarray}

We must fit three parameters ($\lambda$, $\alpha_1$,
$\alpha_2$) using the three measured values
$\Gamma/\Gamma_{TOT}=0.048\pm 0.004$,
$\Gamma_L/\Gamma_T=1.23\pm 0.13$ and
$\Gamma_+/\Gamma_-=0.16\pm0.04$ for the process
$D^+\to\bar{K}^{*0} l^+ \nu_l$,
taken from the Particle Data Group
average of data from different experiments \cite{pdg}.
Unfortunately we are not able to determine
the parameter $\beta$ since $A_0(q^2)$ cannot be observed.

Our model parameters appear linearly in the form factors
(\ref{enfirst})-(\ref{enlast})
and hence in the helicity amplitudes
(\ref{helampone}), (\ref{helamptwo}),
so the polarized decay rates (\ref{defgammapol})
are quadratic functions of them. For this reason
there are $8$ sets of solutions for the three parameters
($\lambda$,$\alpha_1$,$\alpha_2$). It was found
from the analysis of the strong decays $D^*\to D\pi$ and
electromagnetic decays $D^*\to D\gamma$ \cite{ourpaper}, that the
parameter $\lambda$ has the same sign as the parameter
$\lambda'$, which describes the contribution of the
magnetic moment of the heavy (charm) quark. In
the heavy quark limit we have $\lambda'=-1/(6 m_c)$.
Assuming that the finite mass effects are not so large as to
change the sign, we find that $\lambda<0$.
Therefore only four solutions remain. They are shown in
Table \ref{tabthree}.
The errors in Table \ref{tabthree} were calculated from
the experimental errors and the uncertainty in the value
of $f_{Ds*}$ (Table \ref{tabone}).

Of the four possibilities given in Table \ref{tabthree},
only set 2 is acceptable, since it is the only one
to give acceptable values for the form factors at
$q^2=0$. Indeed, for this case we get remarkable agreement
with the experimental data:
$V(0)=1.0 \pm 0.1$ (exp: $1.1 \pm 0.2$),
$A_1(0)=0.57 \pm 0.03$ (exp: $0.56 \pm 0.4$) and
$A_2(0)=0.44 \pm 0.12$ (exp: $0.40 \pm 0.08$).

With these experimentally determined
values of the model parameters it is then
straightforward to calculate the branching
ratios and polarization variables for the other
semileptonic decays of the type $D\to V$. The results
are shown in Table \ref{tabfour}. We see that these
results are in agreement with the known experimental
data.

\vskip 0.5cm

\noindent
{\bf Decays $D\to Pl\nu_l$}

\vskip 0.5cm

There are four Cabibbo allowed semileptonic decays
($D^0\to K^-$, $D^+\to{\bar K}^0$, $D_s^+\to\eta$ and
$D_s^+\to\eta'$) and five Cabibbo suppressed semileptonic
decays ($D^0\to\pi^-$, $D^+\to\pi^0$, $D^+\to\eta$,
$D^+\to\eta'$ and $D_s^+\to K^0$) of the charmed mesons
of the type $D\to P l \nu_l$.

In our approach the form factors are given by

\begin{eqnarray}
\label{ourf1f0}
{1\over K_{HP}} F_1(q^2)=&-&{f_H\over 2}+g f_{H'^*}
{m_{H'^*} \sqrt{m_H m_{H'^*}}\over q^2-m_{H'^*}^2}\;,\\
{1\over K_{HP}} F_0(q^2)=&-&{f_H\over 2}
+g f_{H'^*}\sqrt{m_H\over m_{H'^*}}
(1-2{m_{H'^*}^2\over m_H^2-m_P^2})\nonumber\\
&-&({f_H\over 2}+g f_{H'^*} \sqrt{m_H\over m_{H'^*}})
{q^2\over m_H^2-m_P^2}\nonumber\\
&+&2 g f_{H'^*}(1-{m_{H'^*}^2\over m_H^2-m_P^2})
{m_{H'^*}\sqrt{m_H m_{H'^*}}\over q^2-m_{H'^*}^2}\;.
\end{eqnarray}

\noindent
where the pole masses and the constants $K_{HP}$ are given
in Table \ref{tabfive}.
We shall neglect the lepton mass, so
the form factor $F_0$, which is proportional
to $q^\mu$, does not contribute to the decay width.
The calculation of the decay rate
is very similar to the vector case. After a trivial
integration we obtain

\begin{eqnarray}
\label{defgammap}
\Gamma^P={G_F^2 m_H^2\over
24\pi^3} \int^{y^P_m}_0 dy |F_1 (y)|^2 |\vec{p'}^P(y)|^3\;,
\end{eqnarray}

\noindent
where, similarly as in (\ref{defym})

\begin{equation}
\label{defymp}
y^P_m=(1-{m_P\over m_H})^2\;.
\end{equation}

\noindent
The dimensionless integration variable $y$ has been
introduced with the same definition (\ref{defy}) as
in the vector case and the three-momentum of the
light pseudoscalar meson is given by (\ref{defpp})
with $m_V$ replaced by $m_P$:

\begin{equation}
\label{defppp}
|\vec{p'}^P|^2={[m_H^2 (1-y)+m_P]^2\over 4 m_H^2}-m_P^2\;.
\end{equation}

Using the best known experimental branching ratio -
${\cal B}[D^0\to K^- l^+\nu_l]=(3.68\pm 0.21)\%$
\cite{pdg}, we get two solutions for $g$:

\begin{eqnarray}
\label{solgone}
\hbox{SOL.  1 }&:& g=0.08\pm 0.09\;,\\
\label{solgtwo}
\mbox{SOL.  2 }&:& g=-0.90\pm0.19\;.
\end{eqnarray}

\noindent
The quoted errors are mainly due to the
uncertainties in the value of the heavy
meson decay constant $F_D$.
Unfortunately we are not able to choose between the two
possible solutions for $g$ (\ref{solgone})-(\ref{solgtwo}).
Both solutions give acceptable values for the absolute
value of the form factor at $q^2=0$: $|f_+(0)|=0.87 \pm
0.35$ for solution $1$ and $|f_+(0)|=0.68 \pm 0.34$
for solution $2$, compared to the experimental value
$0.75 \pm 0.03$.

We have calculated the branching ratios for the other
$D\to P$ semileptonic decays, assuming both solutions
for $g$, which give similar results. These results are
summarized in Table \ref{tabsix}.

\vskip 1cm

\centerline{\bf IV. SUMMARY}

\vskip 1cm

We have proposed a method to include the
light vector meson resonances in the
weak currents using HQET and CHPT.
With the use of the weak and strong Lagrangian,
we have analyzed the matrix elements of the
weak currents for the decays
$D^{+}\to {\bar K}^{0*} l^{+}\nu_{l}$
and $D^0\to K^- l^{+}\nu_{l}$.
Instead of the propagators used in HQET we have used the full
propagators for the intermediate heavy meson states.
In this way we obtain a
pole-type behavior of the form factors
for the matrix element of the
vector currents, and a constant behavior
of the form factors in the case of
matrix elements of the axial current.
The unkown parameters $\lambda$, $\alpha_1$ and
$\alpha_2$ were determined using the experimental
measurements of
$\Gamma/\Gamma_{TOT}$, $\Gamma_L/\Gamma_T$ and
$\Gamma_+/\Gamma_-$ for $D^+\to {\bar K}^{*0}$,
giving $\lambda=(-0.34 \pm 0.07)$ GeV$^{-1}$,
$\alpha_1=(-0.14 \pm 0.01)$ GeV$^{1/2}$, and
$\alpha_2=(-0.10 \pm 0.03)$ GeV$^{1/2}$.
{}From the ${\cal B}(D^{0} \to K^{-} l^{+}\nu_{l})$
data the coupling $g$, defined in the strong
Lagrangian for heavy mesons, was determined
as well, but an ambiguity gives two possible
solutions: $g=0.08 \pm 0.09$ and $g=-0.90\pm 0.19$.
We calculated the measured
Cabibbo allowed semileptonic decays
$D^+\to {\bar K}^0$,
$D^+_{s} \to (\eta + \eta')$,
$D^0\to K^{*-}$ and
$D^{+}_{s} \to \Phi$,
and the Cabibbo suppressed decays
$D^{0} \to \pi^{-}$,
$D^{+} \to \pi^{0}$.
The  calculated branching ratios are in agreement
with the experimental results.
We have also predicted the other semileptonic decays,
that have not yet been observed.

\vskip 0.5cm
This work was supported in part by the
Ministry of Science and Technology of the Republic
of Slovenia (B.B. and S.F.) and by the U.S. Department
of Energy, Division of High Energy Physics,
under grant No. DE-FG02-91-ER4086 (R.J.O.).
One of us (B.B.) is very grateful to Lincoln Wolfenstein
and the High Energy Theory Group of Carnegie Mellon
University for the kind hospitality during his stay there,
where part of this work was done.

\begin{table}[h]
\begin{center}
\begin{tabular}{|c|c|c||c|c|c||c|c|}
\hline
$H$ & $m_H$ & $f_H$ &
$P$ & $m_P$ & $f_P$ &
$V$ & $m_V$ \\
\hline
\hline
$D$ & $1.87$ & $0.24 \pm 0.05$ &
$\pi$ & $0.14$ & $0.13$ &
$\rho$ & $0.77$ \\
$D_s$ & $1.97$ & $0.27 \pm 0.05$ &
$K$ & $0.50$ & $0.16$ &
$K^*$ & $0.89$ \\
$D^*$ & $2.01$ & $0.24 \pm 0.05$ &
$\eta$ & $0.55$ & $0.13 \pm 0.008$ &
$\omega$ & $0.78$ \\
$D_s^*$ & $2.11$ & $0.27 \pm 0.05$ &
$\eta'$ & $0.96$ & $0.11 \pm 0.007$ &
$\Phi$ & $1.02$ \\
\hline
\end{tabular}
\end{center}
\caption{\label{tabone}
The pole masses and decay constants in GeV.}
\end{table}

\begin{table}[h]
\begin{center}
\begin{tabular}{|c|c||c|c|c|}
\hline
$H$ & $V$ & $H'^*$ & $H'$ & $K_{HV}$ \\
\hline
\hline
$D^0$ & $K^{*-}$ & $D_s^{*+}$ & $D_s^+$ & $\cos{\theta_c}$ \\
\hline
$D^+$ & ${\bar K}^{*0}$ & $D_s^{*+}$ & $D_s^+$ & $\cos{\theta_c}$ \\
\hline
$D_s^+$ & $\Phi$ & $D_s^{*+}$ & $D_s^+$ & $\cos{\theta_c}$ \\
\hline
\hline
$D^0$ & $\rho^-$ & $D^{*+}$ & $D^+$ & $\sin{\theta_c}$ \\
\hline
$D^+$ & $\rho^0$ & $D^{*+}$ & $D^+$ & $-{1\over\sqrt{2}}\sin{\theta_c}$ \\
\hline
$D^+$ & $\omega$ & $D^{*+}$ & $D^+$ & ${1\over\sqrt{2}}\sin{\theta_c}$ \\
\hline
$D_s^+$ & $K^{*0}$ & $D^{*+}$ & $D^+$ & $\sin{\theta_c}$ \\
\hline
\end{tabular}
\end{center}
\caption{\label{tabtwo}
The pole mesons and the constants $K_{HV}$
for the $D\to V$ Cabibbo allowed and
Cabibbo suppressed semileptonic decays.}
\end{table}

\begin{table}[h]
\begin{center}
\begin{tabular}{|c|c|c|c|}\hline
& $\lambda$ [GeV$^{-1}$]
& $\alpha_1$ [GeV$^{1/2}$]
& $\alpha_2$ [GeV$^{1/2}$]  \\ \hline
SET 1 & $-0.34 \pm 0.07$ & $-0.14 \pm 0.01$ &
$-0.83 \pm 0.04$\\
SET 2 & $-0.34 \pm 0.07$ & $-0.14 \pm 0.01$ &
$-0.10 \pm 0.03$\\
SET 3 & $-0.74 \pm 0.14$ & $-0.064 \pm 0.007$ &
$-0.60 \pm 0.03$\\
SET 4 & $-0.74 \pm 0.14$ & $-0.064 \pm 0.007$ &
$+0.18 \pm 0.03$\\ \hline
\end{tabular}
\caption{\label{tabthree}
Four possible solutions for the model parameters
as determined by the $D^+\to\bar{K}^{*0}l^+\nu_l$ data.}
\end{center}
\end{table}

\begin{table}[h]
\begin{center}
\begin{tabular}{|c||c|c|c|}
\hline
Decay & ${\cal B}$ [$\%$] & $\Gamma_L/\Gamma_T$ & $\Gamma_+/\Gamma_-$ \\
\hline
\hline
$D^0\to K^{*-}$ & $1.8 \pm 0.2$ & $1.23 \pm 0.13$ & $0.16 \pm 0.04$ \\
                & $(2.0 \pm 0.4)$&                 &                 \\
\hline
$D_s^+\to\Phi$ & $1.7 \pm 0.1$ & $1.2 \pm 0.1$ & $0.16 \pm 0.04$ \\
               & $(1.88\pm 0.29)$&$(0.6 \pm 0.2)$ &                 \\
\hline
$D^0\to\rho^-$ & $0.17\pm 0.02$ & $1.34\pm 0.2$ & $0.15\pm 0.10$ \\
\hline
$D^+\to\rho^0$ & $0.22\pm 0.02$ & $1.4\pm 0.2$ & $0.15\pm 0.10$ \\
               & $(<0.37)$      &                               \\
\hline
$D^+\to\omega$ & $0.21\pm 0.02$ & $1.4\pm 0.2$ & $0.16\pm 0.10$ \\
\hline
$D_s^+\to K^{*0}$ & $0.17\pm 0.02$ & $1.3\pm 0.2$ & $0.15\pm 0.10$ \\
\hline
\end{tabular}
\end{center}
\caption{\label{tabfour}
The branching ratios and polarization ratios for the
$D\to V$ semileptonic decays. Where available,
the experimental data is quoted in brackets.}
\end{table}

\begin{table}[h]
\begin{center}
\begin{tabular}{|c|c||c|c|}
\hline
$H$ & $P$ & $H'^*$ & $K_{HP}$ \\
\hline
\hline
$D^0$ & $K^-$ & $D_s^{*+}$ & $(1/f_K)\cos{\theta_c}$ \\
\hline
$D^+$ & ${\bar K}^0$ & $D_s^{*+}$ & $(1/f_K)\cos{\theta_c}$ \\
\hline
$D_s^+$ & $\eta$ & $D_s^{*+}$ &
${1\over\sqrt{8}}[(1-5c^2)/f_\eta-5sc/f_{\eta'}]\cos{\theta_c}$ \\
\hline
$D_s^+$ & $\eta'$ & $D_s^{*+}$ &
${1\over\sqrt{8}}[-5sc/f_\eta+(1-5s^2)/f_{\eta'}]\cos{\theta_c}$ \\
\hline
\hline
$D^0$ & $\pi^-$ & $D^{*+}$ & $(1/f_\pi)\sin{\theta_c}$ \\
\hline
$D^+$ & $\pi^0$ & $D^{*+}$ & $-{1\over\sqrt{2}}(1/f_\pi)\sin{\theta_c}$ \\
\hline
$D^+$ & $\eta$ & $D^{*+}$ &
${1\over\sqrt{8}}[(1+c^2)/f_\eta+sc/f_{\eta'}]\sin{\theta_c}$ \\
\hline
$D^+$ & $\eta'$ & $D^{*+}$ &
${1\over\sqrt{8}}[sc/f_\eta+(1+s^2)/f_{\eta'}]\sin{\theta_c}$ \\
\hline
$D_s^+$ & $K^0$ & $D^{*+}$ & $(1/f_K)\sin{\theta_c}$ \\
\hline
\end{tabular}
\end{center}
\caption{\label{tabfive}
The pole mesons and the constants $K_{HP}$
for the $D\to P$ Cabibbo allowed and
Cabibbo suppressed semileptonic decays.
The $\eta-\eta'$ mixing angle is $\theta_P$ and
$s=\sin{\theta_P}$, $c=\cos{\theta_P}$,
while $\theta_c$ is the Cabibbo angle.}
\end{table}

\begin{table}[h]
\begin{center}
\begin{tabular}{|c||c|c|c|}
\hline
Decay & ${\cal B}_1$ & ${\cal B}_2$ & exp. \\
\hline
\hline
$D^+\to{\bar K}^0$ & $9.4 \pm 0.5$ & $9.4 \pm 0.5$ & $6.7 \pm 0.8$ \\
\hline
$D_s^+\to\eta$ & $3 \pm 3$ & $2 \pm 2$ & \\
\hline
$D_s^+\to\eta'$ & $1.6 \pm 0.7$ & $0.9 \pm 0.5$ & \\
\hline
$D_s^+\to(\eta+\eta')$ & $4 \pm 3$ & $3 \pm 3$ & $7.4 \pm 3.2$ \\
\hline
$D^0\to\pi^-$ & $0.47 \pm 0.05$ & $0.5 \pm 0.5$ & $0.39^{+0.23}_{-0.12}$ \\
\hline
$D^+\to\pi^0$ & $ 0.59 \pm 0.06$ & $0.7 \pm 0.6$ & $0.57 \pm 0.22$ \\
\hline
$D^+\to\eta$ & $0.18 \pm 0.05$ & $0.1 \pm 0.2$ & \\
\hline
$D^+\to\eta'$ & $0.021 \pm 0.005$ & $0.01 \pm 0.01$ & \\
\hline
$D_s^+\to K^0$ & $0.4 \pm 0.2$ & $0.2 \pm 0.3$ & \\
\hline
\end{tabular}
\end{center}
\caption{\label{tabsix}
The branching ratios for the $D\to P$ semileptonic decays,
where ${\cal B}_1$ and ${\cal B}_2$ refer to the two possible
solutions $g=0.08\pm 0.09$ and $g=-0.90\pm 0.19$, respectively.}
\end{table}

\end{document}